# Optical-phonon mediated exciton energy relaxation with highly preserved spin states in a single quantum dot

H. Kumano[1,2], H. Kobayashi[1], S. Ekuni[1], Y. Hayashi[1], M. Jo[1], H. Sasakura[1], S. Adachi[3], S. Muto[3], and I. Suemune[1,2]

[1]*RIES, Hokkaido University, Sapporo 001-0021, Japan*
[2]*Japan Science and Technology Corporation (CREST), Saitama 332-0012, Japan 332-0012, Japan*
[3]*Graduate School of Engineering, Hokkaido University*

High degree of preservation of spin states during energy relaxation processes mediated by optical phonons is demonstrated in a single quantum dot. Optical-phonon resonance and relevant suppression of spin relaxation are clearly identified as dip structures in photoluminescence excitation spectra probed by the positive trion emission. The absence of continuum states makes this observation possible under the cross-circularly polarized detection with respect to a circularly polarized pumping. Consequently, distinguishably high degree of circular polarization up to ~0.85 is achieved without applying external magnetic field at the optical-phonon resonance. Rate equation analysis reveals that the spin-flip probability during energy relaxation is restricted to less than 7.5%. It is also indicated that the spin flip time of the positive trion ground state is extended by more than 3 times compared with that of neutral exciton ground state. This corresponds to the spin flip time longer than 11 ns for the positive trion ground state. The influence of nuclear polarization to the present measurements is also discussed.

Spin states in a single quantum dot (QD) can be one of the most fundamental physical platform for quantum bits (qubits) since spin states in a QD can be much more stabilized than those in higher dimensional systems [1]. They are also promising from the viewpoint of realizing practical solid state devices and their integrations based on highly developed semiconductor technologies. Another essential feature of electronic spin states is the ability to mutually convert into photon polarization states or vise versa via dipole interactions. This enables us to prepare QD-based non-classical Fock-state photons with well-defined polarizations, which are required in quantum key distribution (QKD) [2].

In order to realize efficient state conversions between exciton spins in a QD and the photon polarizations in number states, exciton spins should be highly preserved in each process of the state conversions. Toward this direction, spin relaxation mechanisms in a QD have been widely investigated in terms of the electron-hole (*e-h*) exchange interaction [3,4], the spin-orbit interaction [5,6] as well as the hyperfine interaction [7-10] for both neutral excitons and singly charged excitons (trions). Among them, positively charged trions ($X^+$) are free from the *e-h* exchange interaction [4] and this makes $X^+$ more attractive than neutral excitons ($X^0$). Depolarization of photo-excited electron spins during energy relaxation is generally much slower than hole spins [11], which makes $X^+$ with spin-singlet hole pairs more attractive than negatively charged trions ($X^-$). Furthermore, since the initial and final states of the $X^+$ emission are half spin systems, both states are spin-degenerate due to the Kramer's theorem in the absence of a magnetic field. Thus, the $X^+$ trion state couples to the degenerate two kinds of photons with orthogonal circular polarizations based on the selection rule for the spin angular momentum [12]. Therefore, the $X^+$ trion state is the excellent spin state to examine the state conversions between the exciton spins in a QD and the photon polarizations.

Experimentally, the relaxation of the spin states can be directly examined by measuring the degree of circular polarization (DCP) of photons emitted from a QD under circularly polarized photo-excitation. In actual experimental setups, measurements under the resonant excitation of exciton ground states are rather difficult since the weak photon emission from a single QD is easily masked with the leakage photons from the exciting laser sources. Therefore the spin relaxation during energy relaxation processes is mostly involved in the measured DCP [13-15]. Toward the ideal quantum-state conversions between exciton spin states and photon polarizations, quantitative understanding of both the spin relaxation *during the energy relaxation* and that between the spin-degenerate trion ground states in a single QD is required.

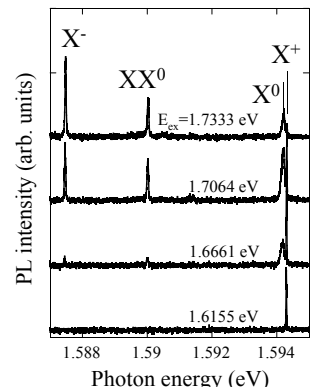

FIG. 1 Excitation energy dependence of the PL spectrum from a single QD. Positively-charged trion ($X^+$), neutral exciton ($X^0$), neutral biexciton ($XX^0$), and negatively-charged trion ($X^-$) are observed.

In this paper, spin flip during the exciton energy relaxation process is studied with detailed measurements of polarization-selective photoluminescence excitation (PLE) spectra and the relevant DCP. The contribution of the optical-phonon resonance for the preservation of the spin states is clearly identified as dip structures in the polarization-selective PLE spectra measured under the cross-circular detection with respect to the excitation polarization. High DCP up to ~0.85 is demonstrated without external magnetic field. The influence of the dynamic nuclear polarization (DNP) is also discussed.

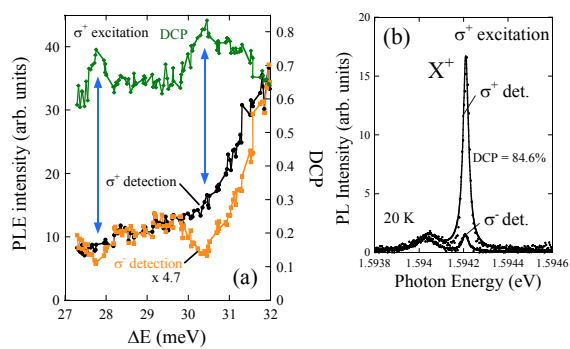

FIG. 2 (a) (color on line) Polarization-selective PLE spectra under $\sigma^+$ excitation. Excitation energy is measured from the $X^+$ emission energy. The PLE spectrum with $\sigma^-$ detection is multiplied by 4.7. Peak structures in DCP and corresponding dip structures in PLE indicated by arrows are attributed to InAs LO (30.5 meV) and TO (27.4 meV) phonon resonances. (b) Polarization-dependent PL spectra under InAs LO phonon-resonant excitation. DCP amounts to ~0.85 for the $X^+$ emission without external magnetic field. The window width for the spin preservation is ~1 meV.

$In_{0.75}Al_{0.25}As$ QDs were employed in this paper and were grown on a semi-insulating (001)-GaAs substrate by molecular-beam epitaxy. The QDs were prepared in Stranski-Krastanow (S-K) growth mode and were sandwiched with $Al_{0.3}Ga_{0.7}As$ layers grown on a semi-insulating GaAs substrate. The topmost surface was terminated with a GaAs cap layer. All the grown layers were nominally undoped, and the details of this sample preparation are described in Ref. 16. After the growth, the sample was etched into mesa structures with diameters of ~150 nm for isolating single QD from the dot ensemble with the density of around $5 \times 10^{10}$ dots/cm$^2$.

For a single dot spectroscopy, the sample was kept at 20 K. A continuous-wave Ti: sapphire laser was used as a circularly polarized excitation source adjusted with a quarter-wave plate. An objective lens with the numerical aperture (NA) of 0.42 focused the laser beam on one of the mesa structures and collected photoluminescence (PL) emitted from the mesa. Collected luminescence was dispersed by a 0.64-m triple monochromator and was introduced into a liquid-nitrogen cooled Si charge-coupled-device detector. The PL polarization was analyzed with $\sigma^+/\sigma^-$ detection employing a set of quarter-wave plate and a fixed Glan-Thomson linear polarizer in front of the monochromator, where $\sigma^+$ ($\sigma^-$) denotes the circular polarization with a helicity of +1 (−1). The overall system resolution was 4.5 μeV.

PL spectra measured from a single QD under the specified excitation energies are shown in Fig. 1. When the excitation energy is above 1.67eV corresponding to the wetting layer (WL) absorption edge, four emission lines originating from $X^+$, $X^0$, $X^-$, and neutral biexciton ($XX^0$) were dominantly observed, which were assigned carefully by several independent measurements [17]. With lowering the excitation energy below that of the WL, excitonic species composed of twin electrons systematically faded away and finally almost exclusive population of the $X^+$ state was observed as exemplified with the excitation energy of 1.6155 eV in Fig. 1. This is because the present QD is populated with a residual hole most probably due to residual acceptors in the $Al_{0.3}Ga_{0.7}As$ layers [18].

Figure 2(a) shows the polarization-selective PLE spectra measured under the $\sigma^+$ excitation and was shown as a function of the excitation energy relative to the detected $X^+$ emission line. Here the energy resolution of the excitation energy was about 50 μeV. The PLE spectrum measured with the $\sigma^-$ detection (hereafter designated as $\sigma^-$-PLE) was multiplied by 4.7 to clearly show the spectral difference to the $\sigma^+$ detection. The DCP defined by $(I^+ - I^-)/(I^+ + I^-)$, where $I^+$ ($I^-$) denotes the PLE intensity measured in the $\sigma^+$ ($\sigma^-$) polarization, is also shown in Fig. 2(a). The multiplication factor of 4.7 given above for the $\sigma^-$-PLE corresponds to the average DCP of 0.65 measured under the non-resonant condition. Two peaks observed in the DCP almost agree with the energies of InAs LO (30.5 meV) and TO (27.4 meV) phonon resonances [19]. The comparison of the PLE spectra with the two cross-polarized detections show that these DCP peaks originate from the reduction of the cross-polarized $\sigma^-$ detection near the LO and TO resonances.

In general, the polarization-selective PLE spectra involve spin flips experienced both *during* and *after* energy relaxation processes to the $X^+$ state. The latter, *i.e.*, spin flips between the two spin-degenerate $X^+$ ground states, is essentially independent of the excitation energy, and the excitation-energy dependence of the polarization-selective PLE spectra predominantly reflects the spin-flips during the energy relaxation. Thus, the dip structures at the LO and TO phonon energies observed in the $\sigma^-$-PLE spectrum are the direct evidence of the suppressed spin-flips during the energy relaxation mediated by the optical phonons. In Fig. 2(a), the increase of the PLE intensities is observed above ~31 meV, which is attributed to the onset of the continuum background transitions in the energy below the WL absorption edge [20]. It is worth mentioning that the observation of the dip structure in the $\sigma^-$-PLE was possible by the critical spectral separation with the continuum background transitions, which will make the spin flip processes much more complex.

Polarization-dependent PL spectra measured under the InAs LO phonon-resonant excitation is shown in Fig. 2(b). Distinguishably high DCP up to ~0.85 is achieved for the $X^+$ emission without external magnetic field. This indicates that the spin polarization is highly stabilized by the InAs LO phonon resonant excitation. The small peak which appeared at the photon energy of 1.5932 eV is the $X^0$ emission and this will be the reflection of the slight influence of the continuum background transition.

In order to elucidate the spin flip in the entire processes involved after photo-excitation, a rate equation analysis was performed. In this model, a quantum dot with a residual hole state $|h^+\rangle$ characterized by the angular momentum projection $J_z=\pm 3/2$ was considered as depicted in the inset of Fig. 3. Preferential virtual excitation of a pair of spin-down electron and spin-up hole by the $\sigma^+$ excitation to the energy one LO-phonon above the $X^+$ state was dealt with, that is, optical-phonon assisted resonant absorption. Doubly degenerate $X^+$ states are characterized by the angular momentum projection $J_z=\pm 1/2$. Then, corresponding rate equations based on a ladder model [21] are given by,

$$\frac{dp_0}{dt} = -\gamma^+ G p_0 - \gamma^- G p_0 + \frac{p_1^+}{\tau_{rad}} + \frac{p_1^-}{\tau_{rad}},$$
$$\frac{dp_1^\pm}{dt} = \gamma^\pm G p_0 - \frac{p_1^\pm}{\tau_{rad}} - \frac{p_1^\pm}{\tau_f} + \frac{p_1^\mp}{\tau_f}, \quad (1)$$

where $p_0$ is the probability of finding the system in the single hole states [22], $p_1^-$ ($p_1^+$) is the probability finding the one in the trion $J_z=-1/2$ ($J_z=+1/2$) state with a spin-down (up) unpaired electron, respectively, $G$ is the excitation rate, $\tau_{rad}$ is the radiative recombination lifetime of the trion, $\tau_f$ is the spin flip time between the $J_z=\pm 1/2$ trion ground states, and $\gamma^+$ ($\gamma^-$) is the spin flip (preservation) probability during energy relaxation under the $\sigma^+$ excitation, which satisfies $\gamma^+ + \gamma^- = 1$.

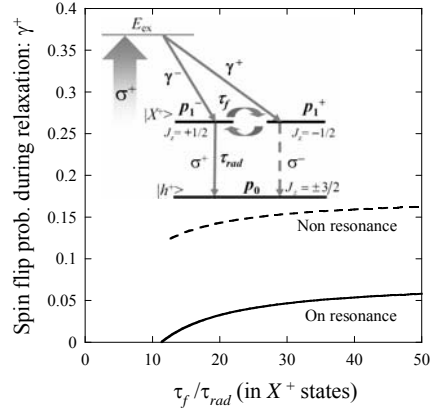

FIG. 3 Relations of two parameters characterizing spin flip during energy relaxation ($\gamma^+$) and in $X^+$ ground states ($\tau_f/\tau_{rad}$) for LO phonon-resonant (solid line) and non-resonant (dashed line) excitations, respectively. DCP value of 0.85 (on resonance) and 0.65 are used for the calculation. Other notations are given in the text. (Inset) Schematic of the model used for rate equation analysis for a QD with a residual hole. Spin flip both *during* and *after* energy relaxation process is included.

By solving the rate equations in steady-state condition, DCP is expressed as

DCP = $(\gamma^+ - \gamma^-)(\tau_f/\tau_{rad})/(\tau_f/\tau_{rad}+2)$. (2)

The relations of $\tau_f/\tau_{rad}$ and $\gamma^+$ which reproduce the experimentally observed DCP of 0.85 and 0.65 under the LO-phonon resonant excitation and non-resonant excitation respectively are given by the solid and dashed lines in Fig. 3. The main points derived from this analysis are twofold: The first point is that the ratio of $\tau_f/\tau_{rad}$ should be larger than 11. Since the radiative lifetime of the $X^+$ state was measured to be 1.0 ns [17], this indicates the spin flip time longer than 11 ns. This ratio of $\tau_f/\tau_{rad}$ is larger than that of the neutral exciton of ~3.6 measured on the identical QD [17]. This is partly because the *e-h* exchange interaction does not work in the trions [4].

The second point is that the allowed range of the spin flip probability $\gamma^+$ during energy relaxation is given from Eq. (2) and Fig. 3. The insertion of infinity for the $\tau_f/\tau_{rad}$ ratio in Eq. (2) gives the maximum spin flip probabilities of 0.075 and 0.175 for the resonant and non-resonant excitations from the measured DCP of 0.85 and 0.65, respectively. Since the allowed $\tau_f/\tau_{rad}$ ratio range is common for the both excitations, the lower limit of $\gamma^+$ for the non-resonant excitation is given as 0.12. Therefore the spin flip probability $\gamma^+$ for the on- and off-optical phonon-resonant excitations is given by the ranges of 0-0.075 and 0.12-0.175, respectively. One can see that the spin flip probability during energy relaxation under the phonon-resonant excitation is substantially suppressed.

In contrast to the $\gamma^+$ parameter, the $\tau_f/\tau_{rad}$ ratio is inherent to the spin-flip mechanisms working on the trion ground states after energy relaxation. Under the steady-state circularly polarized excitation the nuclear magnetic field can be induced by spin-polarized electrons [10], which possibly modifies the $\tau_f/\tau_{rad}$ ratio. Actually, the Zeeman splitting of ~7 μeV was observed in the $X^+$ emission line under the WL excitation (not shown), which corresponds to the magnetic field of $B_N \sim 0.33$ T estimated with the g factor of $g_e = -0.37$ [23]. In order to examine the influence of DNP on the observed high spin stability without an external magnetic field, the polarization modulation measurement of the DCP was performed and the results are shown in Fig. 4. In this experiment, alternate circularly polarized excitation was carried out using an electro-optic modulator [24] and the modulation frequency was swept sequentially. As the modulation frequency was increased, the DCP started to decrease at ~3 kHz and then reaches to the almost constant value above 20 kHz. This behavior was unaffected by the sweep direction. Above 20 kHz where the DNP can no longer follow the modulation, the DCP still exhibits as high as 0.5. This indicates that the DNP has only a subsidiary contribution on the observed high DCP. Concerning the origin of residual spin flip, imperfect initial spin orientation in a photo-absorption process could be partly responsible due to valence band mixing in a QD [25].

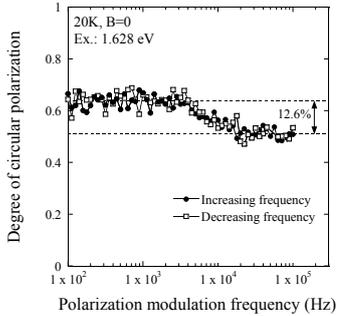

FIG. 4 Polarization modulation frequency dependence of the DCP in $X^+$ line under 1.628 eV excitation. The effect of the DNP on stabilizing the spin state was somewhat limited and estimated to be ~13% in terms of DCP.

In conclusion distinguishably high degree of circular polarization up to ~0.85 was observed in a single quantum dot without external magnetic field by the excitation mediated by optical phonons. Rate equation analysis revealed that the spin flip probability during energy relaxation was highly suppressed to less than ~7% under the optical-phonon resonant excitation. Furthermore, the extension of the spin relaxation time of the positive trion ground states by more than 3 times of the neutral excitons was demonstrated.

The authors would like to acknowledge Dr. H. Z. Song, S. Hirose and M. Takatsu for the sample preparation. The authors are also grateful to Dr. M. Endo for his technical assistance. This work was supported in part by the Grant-in-Aid for Scientific Research (S)(2), No. 16106005, Young Scientists (A), No. 18681025, and Hokkaido Innovation Through Nanotechnology Supports (HINTs) from the Ministry of Education, Culture, Sports, Science and Technology.